\documentclass[sn-basic,Numbered]{sn-jnl}

\usepackage{multicol}

\usepackage{graphicx}%
\usepackage{multirow}%
\usepackage{amsmath,amssymb,amsfonts}%
\usepackage{amsthm}%
\usepackage{mathrsfs}%
\usepackage[title]{appendix}%
\usepackage{xcolor}%
\usepackage{textcomp}%
\usepackage{manyfoot}%
\usepackage{booktabs}%
\usepackage{algorithm}%
\usepackage{algorithmicx}%
\usepackage{algpseudocode}%
\usepackage{listings}%

\newcommand{\eps}{\varepsilon}
\newcommand{\R}{{\mathbb R}} 

\theoremstyle{thmstyleone}%
\theoremstyle{thmstyletwo}%

\theoremstyle{thmstylethree}%

\begin{document}

\title[Information-theoretic evaluation of covariate distributions models]{Information-theoretic evaluation of covariate distributions models}


\author*[1]{\fnm{Niklas} \sur{Hartung}}\email{niklas.hartung@uni-potsdam.de}
\equalcont{These authors contributed equally to this work.}

\author[1,2]{\fnm{Aleksandra} \sur{Khatova}}\email{khatova@b-tu.de}
\equalcont{These authors contributed equally to this work.}

\affil*[1]{\orgdiv{Institute of Mathematics}, \orgname{University of Potsdam}, \orgaddress{\street{Karl-Liebknecht-Str. 24-25}, \city{Potsdam}, \postcode{14476}, \country{Germany}}}

\affil[2]{\orgdiv{Faculty of Health Sciences}, \orgname{BTU Cottbus-Senftenberg}, \orgaddress{\street{Lipezker Straße 47}, \city{Cottbus}, \postcode{03048}, \country{Germany}}}


\abstract{Statistical modelling of covariate distributions allows to generate virtual populations or to impute missing values in a covariate dataset. 
Covariate distributions typically have non-Gaussian margins and show nonlinear correlation structures, which simple descriptions like multivariate Gaussian distributions fail to represent.
Prominent non-Gaussian frameworks for covariate distribution modelling are copula-based models and models based on multiple imputation by chained equations (MICE). 
While both frameworks have already found applications in the life sciences, a systematic investigation of their goodness-of-fit to the theoretical underlying distribution, indicating strengths and weaknesses under different conditions, is still lacking.
To bridge this gap, we thoroughly evaluated covariate distribution models in terms of Kullback-Leibler (KL) divergence, a scale-invariant information-theoretic goodness-of-fit criterion for distributions.
Methodologically, we proposed a new approach to construct confidence intervals for KL divergence by combining nearest neighbour-based KL divergence estimators with subsampling-based uncertainty quantification.
In relevant data sets of different sizes and dimensionalities with both continuous and discrete covariates, non-Gaussian models showed consistent improvements in KL divergence, compared to simpler Gaussian or scale transform approximations.
KL divergence estimates were also robust to the inclusion of latent variables and large fractions of missing values.
While good generalization behaviour to new data could be seen in copula-based models, MICE shows a trend for overfitting and its performance should always be evaluated on separate test data.
Parametric copula models and MICE were found to scale much better with the dimension of the dataset than nonparametric copula models.
These findings corroborate the potential of non-Gaussian models for modelling realistic life science covariate distributions.}




\maketitle

\section{Introduction}

The impact of covariates on an outcome of interest is of relevance in many modelling exercises. 
To name a few examples, age, weight, sex as well as biomarkers and genotypic information enter many pharmacokinetic/pharmacodynamic (PK/PD) models \cite{Hamberg2010,Bajaj2019}, and physiologically-based pharmacokinetic (PBPK) models even rely on a large set of organ weights and blood flows \cite{Jones2013}.
For empirical PK/PD models, a covariate-to-parameter relationship needs to be estimated based on observed data first, which is not necessary for mechanistic PBPK models. 
Irrespective of the model type, the aim then is to perform simulations evaluating the impact of covariates on outcomes of interest.
Often, a set of observed covariates is used for such simulations, but this approach comes with important limitations: 
(i) Mostly, only summary statistics are shared --- other researchers cannot reproduce the results since the information on the correlation structure is lost; (ii) Specific subpopulations may be too sparsely observed for meaningful simulations; (iii) Missing values in a covariate dataset need to be dealt with.
In contrast, statistical models for covariate distributions allow to generate virtual populations of arbitrary sizes and can often naturally deal with missing values.
Since covariate distributions typically have non-Gaussian margins and show nonlinear correlation structures, simple approaches based on multivariate Gaussian distributions may fail to represent them accurately.

A natural approach to overcome a Gaussian assumption for margins of continuous covariates is to apply a scale transform (e.g., log-transform for intrinsically positive covariates).
An effort to include discrete variables into such a lognormal model has been made by \cite{Tannenbaum2006}. 
Recently, two non-Gaussian methods, namely Multivariate Imputation by Chained Equations (MICE) and copula-based models, have also been proposed for the simulation of covariate distributions.
MICE, an imputation method for missing data \cite{Buuren2011}, has been applied for the simulation of covariate distributions in \cite{Smania2021}.
Copula modeling, which has a history in financial mathematics, has been introduced to life sciences by \cite{Zwep2023}.
Beyond these data-driven approaches, mechanistic scaling methods have been proposed for some domains like PBPK model parametrization (organ weights/blood flows) \cite{Huisinga2012}.
Also, regression models have been developed for some examples, like the distribution of body weight as a function of age and sex \cite{Sumpter2011}.
Despite these efforts, multivariate covariate distribution modelling remains the most generally applicable method.

Existing studies have evaluated model performance in terms of univariate or linear measures of association (mean and (co-)variance), which do not discriminate between Gaussian and non-Gaussian models, or by visual inspection \cite{Smania2021,Zwep2023}.
However, a goodness-of-fit metric for the evaluation of covariate distribution models needs to be able to detect \textit{any} deviation, including non-Gaussian, from the data generating distribution.
Kullback-Leibler (KL) divergence fulfils these demands; furthermore it is scale-invariant and can be interpreted well, namely as the information loss when approximating the true population distribution by a surrogate model \cite{Kullback1951}.
Also, typical complicating factors like mixed discrete/continuous data, missing data, latent variables and overfitting have received little attention. 
We therefore aimed at evaluating goodness-of-fit via KL divergence and in a variety of situations, also covering the above-mentioned issues.

Since in practice, only a sample from the data generating distribution is available, sample-based estimation of KL divergence is required.
For high-dimensional continuous data, methods based on nearest neighbour density estimation, combined with a finite sample bias correction, have been found to perform well \cite{PerezCruz2008,Wang2009}.
However, both the extension to mixed discrete/continuous data as well as uncertainty quantification have not been discussed in the literature. 
Therefore, a secondary goal of the present work is to extend the available methodology to accommodate mixed data types (a common case in covariate data sets) and to determine confidence intervals for KL divergence estimates (which allows to judge the significance of differences in KL divergence estimates between models/scenarios).
An implementation of this approach, combining KL divergence estimators with subsampling-based uncertainty quantification, has recently been distributed as \textbf{R} package \texttt{kldest} \cite{kldest-Package}.

\section{Methods}

\subsection*{Datasets}

We considered three different datasets that cover both data-rich and -sparse, as a well low- and high-dimensional situations, thereby providing a broad range for the benchmarking covariate distribution models.

\begin{description}
\item[NHANES-3/-11.] The US national health and nutrition examination survey (NHANES) contains a collection of demographic, physical, health-related and lifestyle-related variables from a representative sample of the US population \cite{NHANES-Data}. 
The dataset distributed in \textbf{R} package \texttt{NHANES} contains a resampled version of $n=10000$ individuals from the 2009-2012 NHANES survey, to counterbalance oversampling of certain subpopulations \cite{NHANES-Package}. 
From this dataset, we ignored individuals with age of 80 years or above (for whom exact age is not given) and those with a diastolic blood pressure of 0 (likely a database error). Subsequently, we derived two NHANES analysis datasets by selecting a subset of covariates and removing duplicates and missing values:
\begin{description}
\item[NHANES-3] (data-rich low-dimensional setting). 6230 unique measurements of 3 continuous covariates: age, weight and height. 
\item[NHANES-11] (medium-sparse high-dimensional setting). 2133 unique measurements of 11 covariates, of which 10 are continuous (age, weight, height, heart rate, systolic blood pressure, diastolic blood pressure, testosterone concentrations, total cholesterol, urine volume, urine flow rate) and one is discrete (sex).  
\end{description}

\item[ORGANWT] (very sparse high-dimensional setting). A dataset on organ weights of two different mouse strains has been reported by \cite{Marxfeld2019}. From this dataset, we selected 6 continuous covariates (body weight and the 5 most relevant organs weights for PBPK modelling \cite{Jones2013} that are available in the dataset, namely brain, heart, kidneys, liver, spleen) and 2 discrete variables (sex and strain), totalling 145 measurements of 8 covariates (after removing one animal in which brain weight was missing).
\end{description}

The sizes of the datasets are summarised in Table~\ref{tab:datasets}.

\begin{table}
\begin{tabular}{c|cc|c}
  & \multicolumn{2} {c|} {Number of covariates} & Number of \\
  & continuous & discrete & measurements \\
\hline
NHANES-3 & 3 & 0 & 6230 \\
NHANES-11 & 10 & 1 & 2133 \\
ORGANWT & 6 & 2 & 145 \\
\hline
\end{tabular}
\caption{Dimensions of the three datasets.}\label{tab:datasets}

\end{table}

\subsection*{General notation and scope}

We consider the observed covariates $\mathbf{x} = (x^{(1)},...,x^{(n)})$
to be an i.i.d.~sample from an (unknown) $d$-dimensional probability distribution with density $p$ and (cumulative) distribution function $F$. For example, in dataset NHANES-3, $x^{(i)}=(x^{(i)}_{1},x^{(i)}_{2},x^{(i)}_{3})$ is age, weight and height of the $i$-th observed individual ($d=3$). 
The aim of covariate distribution modelling is to estimate a surrogate model $q$ which approximates $p$,
in order to be able to simulate i.i.d.~data $\mathbf{y}$ from this surrogate model. Alternatively, a method may directly provide such data $\mathbf{y}$.

\subsection*{Copula models}

Here we give a short introduction to copula modelling for continuous distributions.
Since the notation required for a general presentation of theory and estimation (especially for vine copulas) is complex, we refer to \cite{Czado2019} for further details. 
Any multivariate density $p$ can be uniquely decomposed into a product of marginals and their dependency structure (known as Sklar's theorem),
\begin{equation}
\label{eq:copula-decomposition}
p(x_{1},...,x_{d}) = p_{1}(x_{1})\cdot\ldots\cdot p_{d}(x_{d})\cdot c\big(F_{1}(x_{1}),...,F_{d}(x_{d})\big),
\end{equation}
where $p_{i}$ and $F_{i}$ are the $i$-th marginal density and distribution functions, respectively, and where $c$ is a copula density, supported on the unit cube.
This decomposition allows to estimate the marginals and the dependency structure in two separate steps. 

First, the marginals are modelled, giving rise to estimated marginal distribution functions $\hat F_{1}, ..., \hat F_{d}$. In this step, standard univariate distribution fitting methods can be used.
Here, we employed univariate local polynomial kernel density estimators as implemented in \textbf{R} package \texttt{kde1d} \cite{kde1d-Package}. If the margins are regular enough, parametric methods can be used as well.

Next, using the estimated marginals, the covariates are transformed to uniform scale via 
\begin{equation}
\label{eq:scale-transform}
x \mapsto \hat u(x) := \big(\hat F_{1}(x_{1}),...,\hat F_{d}(x_{d})\big).
\end{equation}
A copula model can then be estimated on the transformed data $\mathbf{u} = \hat u(\mathbf{x})$. One possible approach consists of using multivariate copula models such as Gaussian copulas. While this approach is conceptually simple, it assumes the same type of dependency across all variables. A more flexible copula modelling approach consists of combining bivariate copula models in a pair copula decomposition of the multivariate copula. These copula models, termed vine copulas, use a graph-theoretical structure (vine) describing the conditional dependency structure of the pair copula decomposition.
Modelling with vine copulas is supported by libraries such as \texttt{rvinecopulib} \cite{rvinecopulib-Package}, which automatize the selection of both the vine structure and the pair copulas, and allow for efficient simulation.

\subsection*{Multiple imputation by chained equations (MICE)}

MICE is, at its origin, a method for imputing missing values in a datasets. Also known as conditional distribution modelling, it aims at predicting a missing value given all other (non-missing) observations. Within MICE, different algorithms are available. We used the defaults in the \textbf{R} implementation in package \texttt{mice} namely predictive mean matching \cite{Little1988} for continuous variables and logistic regression for categorical variables, for which robust performance has been reported \cite{Buuren2011}. To use this method for the simulation of covariate distributions, we used the method described in \cite{Smania2021}: (i) if missing data are present, a single standard MICE step is used to fill the missing data; (ii) then, all observed covariates are again labeled as missing and drawn in turn according to MICE in order to obtain a sample.

\subsection*{Covariate distribution models}

Six different covariate distribution models are considered in this article, which differ in how marginals are modelled, the choice of dependency structure and the use of parametric or nonparametric elements. 

\begin{description}
\item [GaussDist.] A multivariate Gaussian distribution (with density $q$) is fitted to the data $\mathbf{x}$ on original scale. Hence, all marginals are also Gaussian. This method only applies to continuous covariates.
\item [IndepCop.] An independence copula is used for the transformed data $\mathbf{u}$. This means that marginals are modelled, but no dependency structure.
\item [GaussCop.] A multivariate Gaussian copula is fitted to the transformed data $\mathbf{u}$. This scenario combines \emph{GaussDist} and \emph{IndepCop}. It can be considered an optimized version of the ``continuous method'' by \cite{Tannenbaum2006}, in which the log-transform is replaced by an kernel-based transform.
\item [ParVine.] A vine copula is fitted to the transformed data $\mathbf{u}$. For each dependency element within the vine copula structure, the best bivariate copula is selected from a family of parametric copulas.
\item [NonparVine.] A vine copula is fitted to the transformed data $\mathbf{u}$. For each dependency element, a nonparametric bivariate copula model is constructed based on a transform kernel approach \cite{Nagler2017}.
\item [MICE.] The MICE method is applied to the entire dataset, as described in \cite{Smania2021}. 
\end{description}

\subsection*{Kullback-Leibler divergence estimation}

For ease of notation, we assume here that all considered covariates are continuous. A combined continuous/discrete framework is described in Appendix~\ref{sec:combined-framework}.
We assume a surrogate model with density $q$ (i.e., a multivariate Gaussian distribution or a copula model) has been estimated from the data $\mathbf{x}$.
The KL divergence between $p$ and $q$ is then defined as 
\begin{equation}
\label{eq:KLD}
D_\text{KL}(p||q) := \int_{\R^{d}} \log\left(\frac{p(x)}{q(x)}\right)p(x)dx.
\end{equation}
To estimate $D_\text{KL}(p||q)$, we used a nearest neighbour density-based method proposed in \cite{Wang2009}. 
In their approach, $D_\text{KL}(p||q)$ is estimated from two samples, $\mathbf{x}$ from $p$ and $\mathbf{y}$ from $q$.
While $\mathbf{x}$ is already available, a (large) sample $\mathbf{y} = (y^{(1)},...,y^{(m)})$ is simulated from the surrogate model $q$. For copula-based models, $\mathbf{y}$ is obtained by transforming a copula sample $\mathbf{v}$ back to the original scale, $\mathbf{y} = \hat u^{-1}(\mathbf{v})$. 
Working with a second sample $\mathbf{y}$ rather than directly using the surrogate model density $q$ allowed for the use of a two-sample bias reduction method. 
Also, it allows a natural comparison to the MICE method, in which a sample $\mathbf{y}$ is obtained directly without estimating a density first. 

The construction of the two-sample bias-reduced nearest-neighbour KL divergence estimator is presented in Appendix~\ref{sec:kld-estimator}. 
It is implemented and documented in the \textbf{R} package \texttt{kldest} \cite{kldest-Package}.
A benchmark (for different data dimensions and with uncertainty quantification) demonstrating robust performance of this estimator is presented on the website accompanying the \textbf{R} package \texttt{kldest} \cite{kldest-Website}.

While KL divergence is scale-invariant, KL divergence \textit{estimation} might differ between data scales. 
In the main text, all results are shown on the original scale. 
An investigation of differences between original and uniform scale (where the copula models are fitted) is shown in Figure~\ref{fig:scale-dependency} (Appendix~\ref{sec:scale-dependency}).

\subsection*{Uncertainty quantification}

Uncertainty quantification of KL divergence estimators has not been addressed in the literature beyond the 1-D discrete case, and in particular, no asymptotic distributions of nearest neighbour-based KL divergence estimators are available.
Also, nearest neighbour-based KL divergence estimation cannot deal with samples containing duplicates. 
This means that standard bootstrapping, which relies on resampling with replacement, cannot be used for uncertainty quantification either. 
Instead, we opted for using subsampling as described by \cite{Politis1994}. 
A precise definition of the procedure is provided in Appendix~\ref{sec:subsampling}. 
Briefly, a large number $s$ of subsamples of size $b \ll n,m$ is drawn without replacement. 
Subsequently, the distribution of the KL divergence estimator is approximated by the subsample distribution, corrected for differences in sample size. 
We chose $b = n^{2/3}$ and large fixed $s = 1000$ and $m=10000$.

\subsection*{Simulation framework}

All simulations were performed in R version 4.2.2 \cite{R-Software}. The source code for reproducing the results, including the three datasets, is available on Zenodo \cite{Hartung2024}.
Requiring as input a list of datasets, it is designed to easily generalise to other datasets. 

\section{Results}

\subsection*{Non-Gaussian methods lead to a considerably better fit}
\begin{figure}[htp!]
\centering
\includegraphics[width=.32\textwidth]{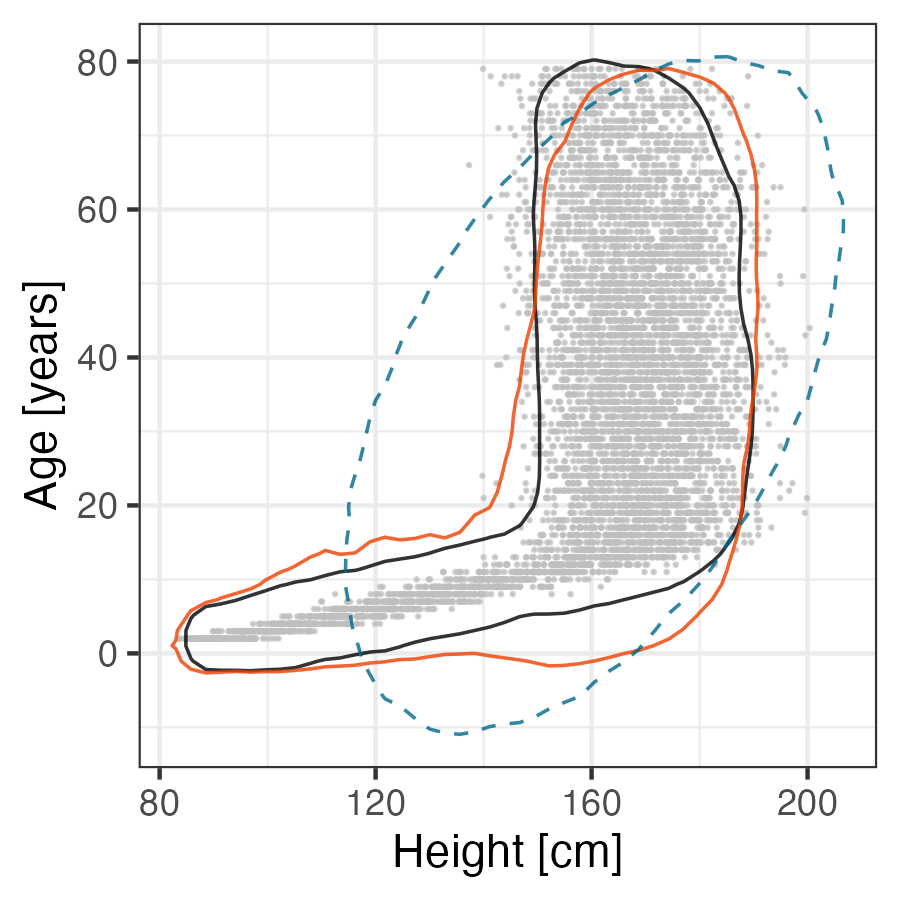}
\includegraphics[width=.32\textwidth]{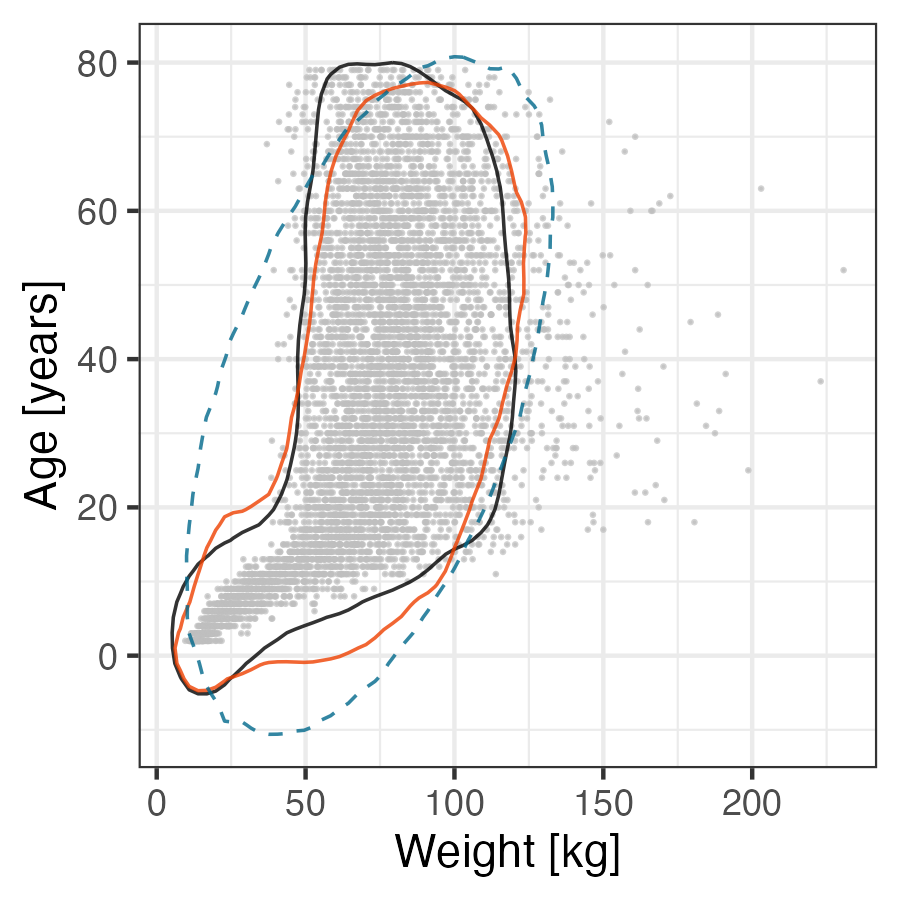}
\includegraphics[width=.32\textwidth]{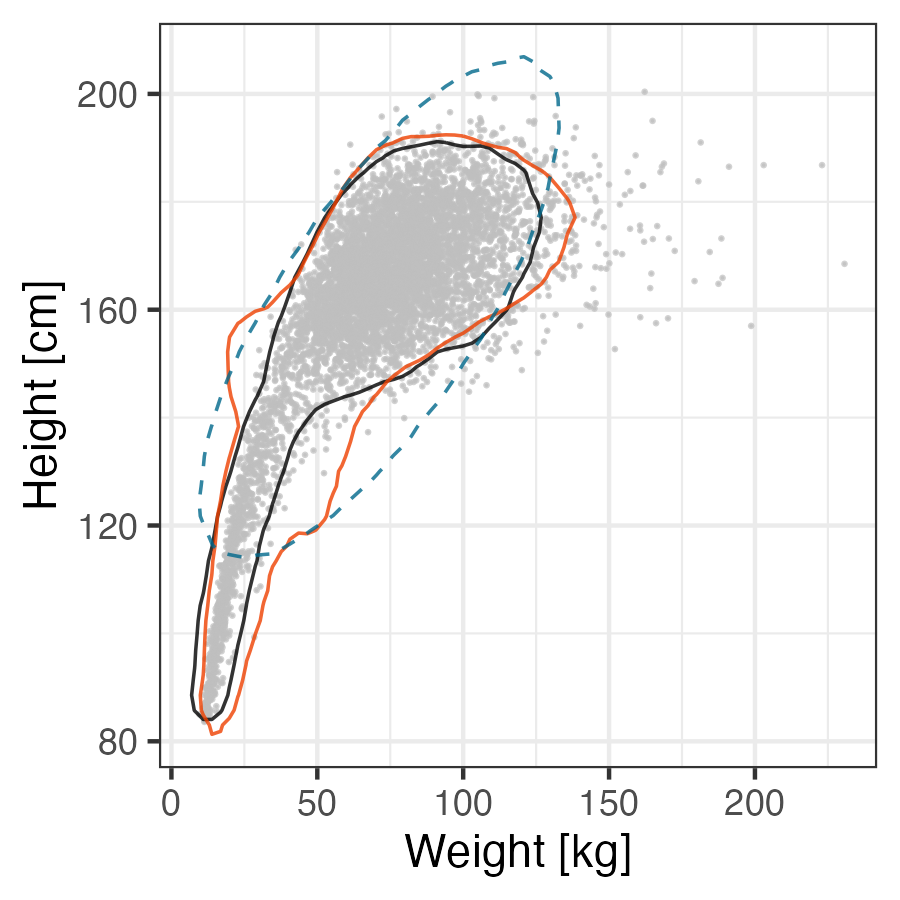}
\includegraphics[width=.8\textwidth]{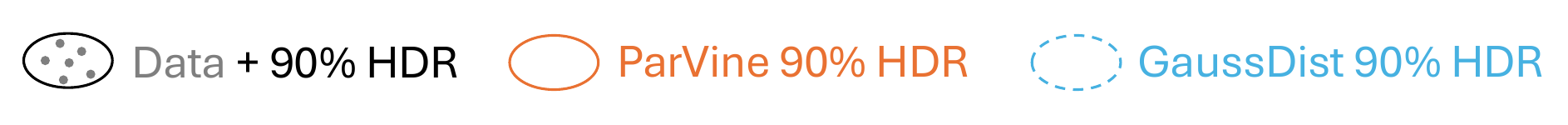}
\caption{
\label{fig:fit-age-weight-height}
Scatterplot matrix and 90\% highest density regions (HDR), obtained by bivariate kernel density estimation, for dataset NHANES-3 and two fitted covariate distribution models, ParVine and GaussDist.
}
\end{figure}

We start by considering dataset NHANES-3. As can be seen in Fig.~\ref{fig:fit-age-weight-height}, the three covariates age, weight and height have a non-Gaussian dependency structure, which a parametric vine copula model (ParVine) is able to capture.  
Importantly, the clear visual improvement in goodness-of-fit of ParVine over GaussDist is \emph{not} reflected in linear metrics: the mean and (co-)variance of GaussDist coincides with the respective empirical metric in the dataset (e.g., mean age $\pm$ SD $\approx35.0\pm21.1$ years), while that of ParVine differs somewhat (mean age$\pm$ SD $\approx32.3\pm21.0$ years). In contrast, KL divergence is much lower for ParVine than for GaussDist (see also Table~\ref{tab:kld}), as would be expected from graphics.

Similar non-Gaussian relationships are present in the higher-dimensional datasets NHANES-11 and ORGANWT, see  Figures~\ref{fig:pairs-nhanes11} and \ref{fig:pairs-organwt} (Appendix~\ref{sec:supp-figs}).
KL divergence estimates for all three datasets are summarized in Table~\ref{tab:kld}. For the large and low-dimensional dataset NHANES-3, MICE performed best, followed by NonparVine and ParVine. In the high-dimensional dataset NHANES-11, MICE still had the best performance, but ParVine was better than NonparVine, showing that the two vine copula-based methods scale quite differently with the dataset dimension. Lastly, in the sparse dataset ORGANWT, ParVine showed better performance than MICE and NonparVine. In all three datasets, the two vine copula-based methods and MICE outperformed the simpler (e.g. Gaussian) approaches.

\begin{table}
\begin{tabular}{c|ccc}
  & NHANES-3 & NHANES-11 & ORGANWT\\
\hline
GaussDist & 1.12 (1.11--1.25) & -- & --\\
IndepCop & 1.30 (1.27--1.42) & 1.72 (1.63--2.04) & 3.62 (2.91--4.92)\\
GaussCop & 0.99 (0.97--1.12) & 1.45 (1.39--1.75) & 2.73 (2.00--3.83)\\
ParVine & 0.56 (0.52--0.66) & 0.62 (0.51--0.85) & \textbf{1.96 (1.16--2.77)}\\
NonparVine & 0.46 (0.42--0.57) & 1.03 (0.95--1.27) & 2.33 (1.61--3.24)\\
MICE &  \textbf{0.21 (0.17--0.30)} & \textbf{0.36 (0.25--0.58)} & 1.96 (1.77--3.27)\\
\hline
\end{tabular}
\caption{
Estimated Kullback-Leibler divergence (with 95\% CI) for all 6 covariate models and 3 datasets. The lowest  KL divergence estimate per dataset is highlighted in bold.
}\label{tab:kld}
\end{table}

\subsection*{MICE, but not copula models, shows overfitting to the data}

\begin{figure}[htp!]
\centering
\includegraphics[width=\textwidth]{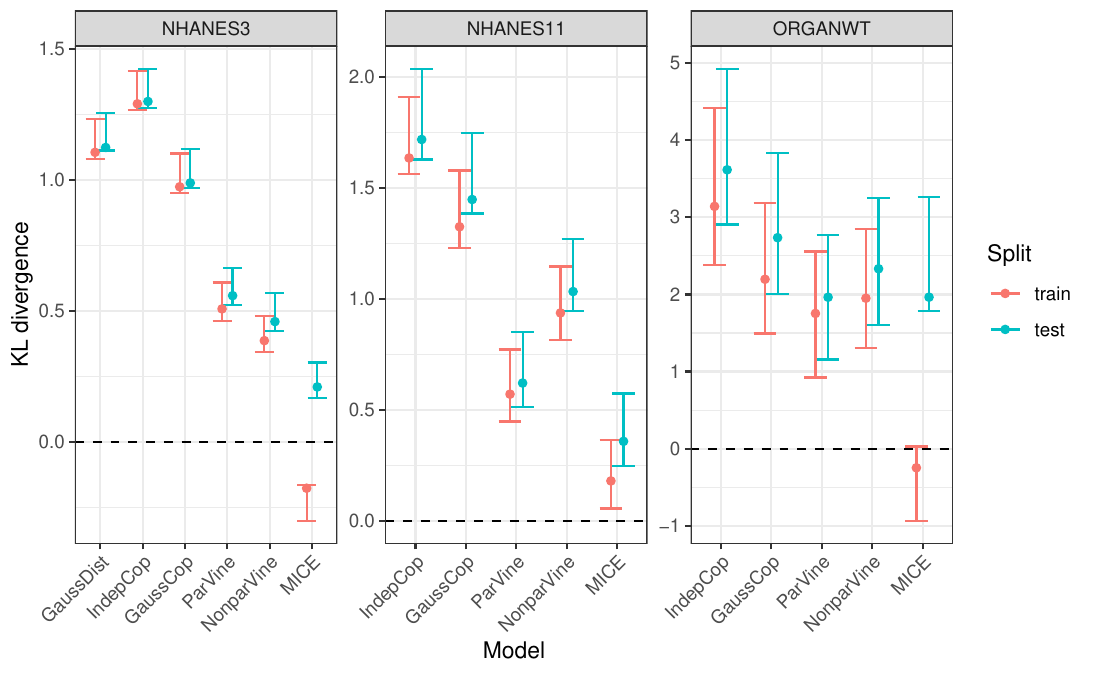}
\caption{
\label{fig:training-vs-test}
Comparison of KL divergence estimates (with 95\% CI) between training and test data, for all considered covariate distribution models and datasets.
}
\end{figure}

Due to the flexibility in the shapes that non-Gaussian covariate distribution models can describe, the question naturally arises whether they tend to overfit the data. We investigate this question by splitting the NHANES-3, NHANES-11 and ORGANWT dataset in half at random, fitting models on one half and comparing KL divergence to the fitted (training) vs. the non-fitted (test) data.
As shown in Fig.~\ref{fig:training-vs-test}, Gaussian and copula-based models have very similar KL divergence estimates on the two data splits. In contrast, the MICE method tends to overfit the data, with much lower KL divergence estimates on the training compared to the test data. To account for this difference in model behaviour, all simulations results shown in this article employ this data split and evaluate KL divergence on the test data only. While Fig.~\ref{fig:training-vs-test} only shows one data split, choosing other splits does not qualitatively change the results.
Fig.~\ref{fig:training-vs-test} also illustrates that the estimator can take negative values. In practice, this would be interpreted as 0.

\subsection*{All models tolerate a moderate degree of missing data}

\begin{figure}[htp!]
\centering
\includegraphics[width=\textwidth]{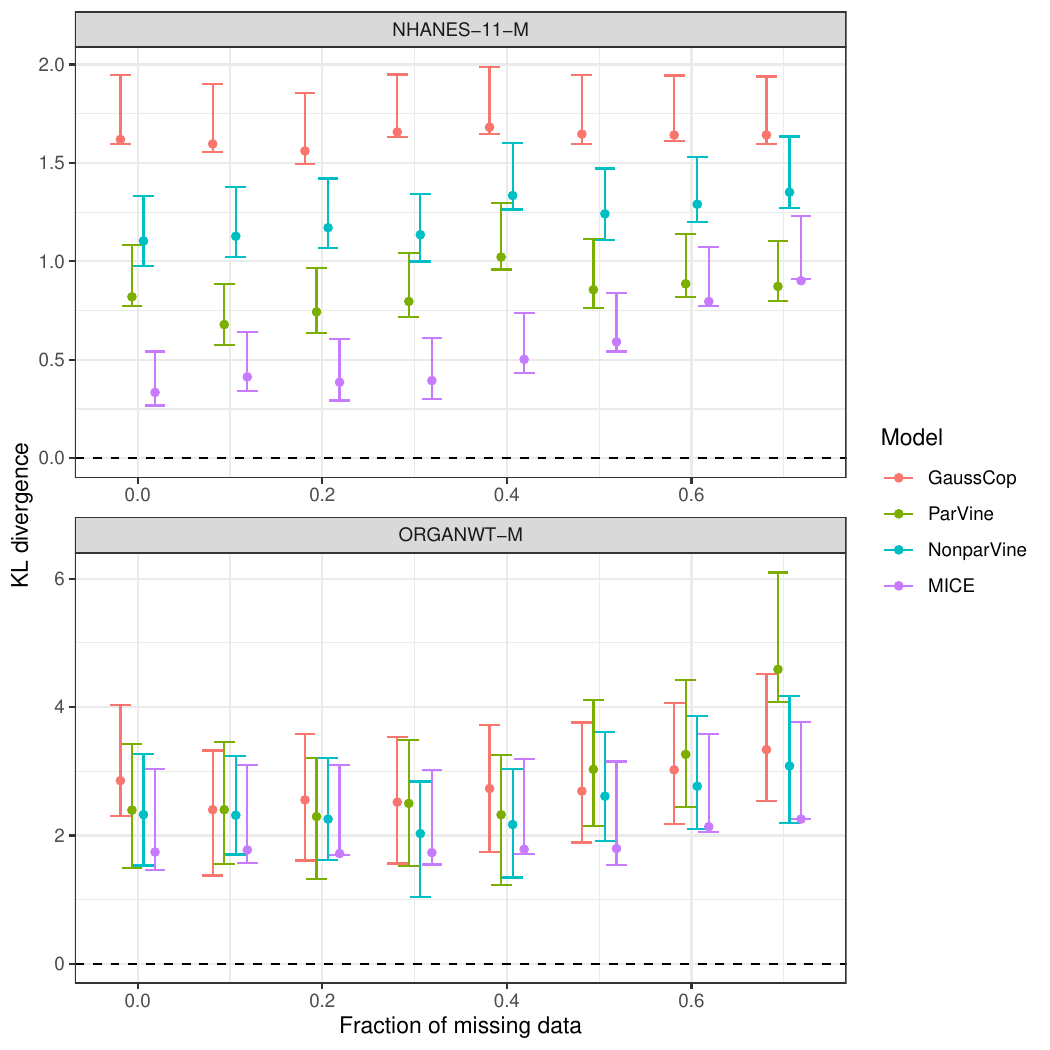}
\caption{
\label{fig:kld-missing-data}
Comparison of KL divergence estimates (with 95\% CI) for different fractions of missing data, for copula-based covariate distribution models (excluding IndepVine) and MICE, and two data scenarios. 
}
\end{figure}

Both the copula framework and MICE can handle missing data, and we investigated the fit to the data for different degrees of missingness. Methods GaussDist and IndepCop were excluded from this analysis since GaussDist would need additional post-processing and IndepCop assumes covariates to be independent, which prohibits useful inference for missing values from the model. 
Two scenarios were considered: 

\begin{itemize}
\item NHANES-11-M: datasets with different fractions $p$ of data missing completely at random were simulated. Of note, due to the high dimensionality of this dataset, the fraction of complete observations $f_\text{complete} = (1-p)^{11}$ quickly decreases with $p$ (e.g., $f_\text{complete} < 0.02$ for $p=0.3$).
\item ORGANWT-M: datasets with complete observations on sex, strain and body weight, but fractions $p$ of missing organ weights are simulated. This corresponds to the common case where some covariates are much easier to measure than others. 
\end{itemize}
Subsequently, covariate distribution models were fitted to the incomplete datasets, samples were generated from the estimated covariate distributions and KL divergence estimated to an independent (complete) test dataset. The results of this analysis are depicted in Fig.~\ref{fig:kld-missing-data}. All investigated methods were able to handle large fractions of missing data, with almost unchanged performance for up to $p=0.3$ and mostly moderate increases in KL divergence thereafter. 

\subsection*{Latent variables do not compromise model performance}

\begin{figure}[htp!]
\centering
\includegraphics[width=\textwidth]{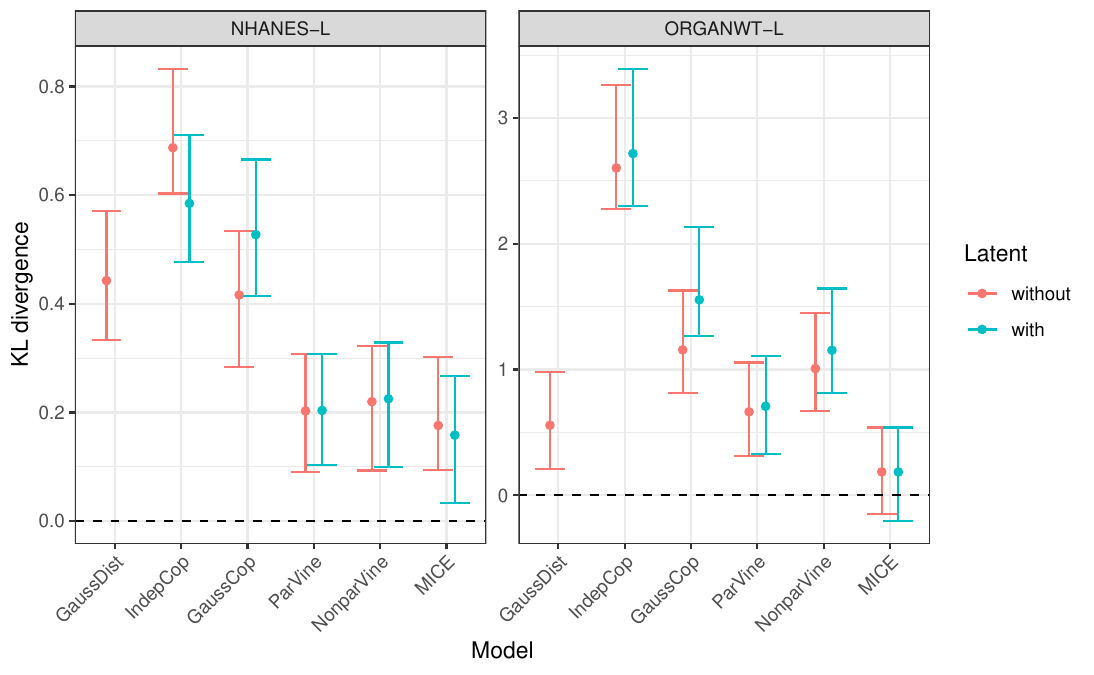}
\caption{
\label{fig:kld-latent-variables}
Impact of latent variables in covariate distribution models on  KL divergence estimates (with 95\% CI). Since the multivariate Gaussian distribution model (GaussDist) can deal with the observed continuous covariates, but not with the discrete latent variables present in both datasets, the second error bar is missing for model GaussDist. 
}
\end{figure}

The variables included during covariate distribution model development may not all be needed in a given application scenario. 
Thus, the question naturally arises how fitting a high-dimensional covariate distribution impacts on the fit to a subset of these covariates. 
To formally address this question, we split a covariate vector $x$ into observed and latent covariates, $x = (x_{o},x_{l})$, and investigate the KL divergence between the distribution of observed covariates and the joint covariate distribution model, subsequently marginalized over the latent covariates. 
We considered two scenarios: 
\begin{itemize}
\item  NHANES-L: impact of including all variables in NHANES-11 while only those in NHANES-3 are of interest.
\item  ORGANWT-L: impact of including body weight, sex and strain as latent variables when the variables of interest are the organ weights.
\end{itemize}
In both scenarios, the inclusion of latent variables only had a minor impact on estimated KL divergence (Fig.~\ref{fig:kld-latent-variables}). 
Therefore, including additional covariates during development of a covariate model allows to increase its range of applicability without deteriorating model performance.

\section{Discussion}

In this work, the performance of non-Gaussian covariate distribution models was evaluated systematically by means of KL divergence between the underlying distribution and the estimated covariate distribution model. 
Non-Gaussian methods showed considerably improved goodness-of-fit over Gaussian ones, and packages are available that greatly facilitate the use of these more advanced methods.

Parametric vine copulas (ParVine) and MICE performed best overall. 
When deciding between these two approaches, two main points need to be considered. 
First, MICE generally works better than ParVine in a rich data scenario, while ParVine performs as good as MICE for sparse data. The second aspect concerns data sharing: while parametric model fits can be shared easily (e.g., as a table) without having to disclose original data, MICE always relies on the underlying dataset. Therefore, if a dataset cannot be shared, e.g. for intellectual property reasons, a parametric vine copula model like ParVine still allows for a reproducible analysis and easier adoption by other researchers.
The nonparametric vine copula model did not perform as well as one might have expected. While it performed better than ParVine in the low-dimensional and observation-rich dataset NHANES-3, the other nonparametric method, MICE, was even better. Also, increasing dimensionality affected the nonparametric vine copula method much more than the other methods. 

All copula-based models and MICE were robust to quite large fractions of missing values and inclusion of latent variables. Therefore, building a high-dimensional and/or sparse dataset for covariate modelling does not invalidate the use of such models on a subset of covariates. Since different PK/PD or PBPK models require different covariates as input and complete observations may be difficult to obtain, this is a desirable feature.

The MICE methodology is quite different to the other (distribution-based) models, which has several implications. First, MICE may produce (generally, a small fraction) of simulated values that are identical to observed sets of covariates since the imputed values are drawn from the set of observed values. In this respect, MICE behaves like a (more refined) bootstrapping procedure. Since nearest neighbour-based KL divergence estimator cannot deal with ties in the data, these were removed from the simulated covariate set. Also, sampling from observed values makes MICE much more prone to overfitting, which needs to be accounted for when comparing different models. Here, this was realized by a training/test data split approach.

KL divergence was chosen as a measure to quantify the information loss in an estimated covariate model compared to the data-generating process. While this measure is frequently used and has good theoretical properties, alternatives do exist. For example, other types of $f$-divergences, such as Hellinger distance or total variation distance, could be used \cite{Renyi1961}. An alternative of a different type would be the Wasserstein distance \cite{Kantorovich1960}, but it is not scale-invariant.

Sample-based KL divergence estimation, including uncertainty quantification and mixed continuous/discrete data, was rendered possible by combining nearest neighbour density estimation with subsampling and a reformulation conditional on the discrete covariates.
While this approach extends already existing methods significantly, some caveats do apply. 
First, the conditional formulation does not scale well with the number of categories and discrete variables.
Also, KL divergence estimation by nearest neighbour-based methods cannot deal with ties in the dataset. 
Such ties could appear at several levels, namely the data itself (which might be resampled for representativity reasons, as the NHANES data), the covariate distribution method (as it may happen in MICE), or resampling-based uncertainty quantification. 
In this work, all ties were avoided, either by the choice of method (subsampling instead of bootstrapping) or as a post-processing step (for NHANES and MICE). 
Jittering could be an alternative to this approach, but the jittering strength is another parameter which is expected to have a significant impact. 
Therefore, this option was not considered in the present work.

While the investigated covariate distribution models cover a broad spectrum of modelling practice, some additional approaches are worth mentioning. 
First, bootstrapping from the covariate set $\mathbf{x}$ is often done \cite{Teutonico2015}. 
Since all samples $y^{(i)}$ produced this way have an identical counterpart in the dataset $\mathbf{x}$, the KL divergence estimator cannot deal with this method -- again, jittering would be an option. 
Although bootstrapping was excluded from our analysis, we did consider MICE, which has many similarities with bootstrapping (and, in addition, can readily deal with missing data). 
Second, generative machine learning models like diffusion models have been applied with success to the simulation of, e.g., realistic images \cite{Cao2024}. 
In principle, such methods could also be applied to covariate distribution modelling, but since covariate datasets are differently structured (non-spatial) and much more sparse when compared to image data, considerable adjustments would certainly be required for these methods, which were out of scope of this work.

The considered datasets were chosen based on relevancy of the considered covariates. 
Covariates age, weight, height and sex are among the most commonly used ones in empirical PK models, and are also used as input parameters in physiological scaling methods \cite{Huisinga2012}. 
The mouse organ weight dataset is of high relevance for PBPK modelling, a context in which a detailed physiology is required. 
These datasets covered both sparse and rich data scenarios and varying dimensions. 
Importantly, the training/test data split ensured that the predictive behaviour of covariate distribution models was evaluated, not a mere fit. 
We expect our results to generalise to other covariate datasets consisting mainly of continuous variables (as is often the case in PK/PD modelling) and with similar context and dimensions as the considered ones. 
However, since a KL-based evaluation becomes more challenging in the presence of both discrete and continuous covariates, only a limited number of discrete (binary) covariates were considered. 
Consequently, it is less clear whether the analyses generalise to datasets with many discrete variables or categories.

Furthermore, all covariates were considered static, i.e. time-varying covariates were out of scope. First results on copula modelling for time-varying covariates were given in \cite{Zwep2023}. Also, the impact of covariate distribution model misspecification on PK/PD or PBPK model predictions was not investigated. While the question is of much relevance and could be investigated in future work, we decided to focus on the covariate distribution model itself, in order to sharpen the focus and not render the presented analysis dependent on another model type (a specific PK/PD or PBPK model). 
A correctly specified covariate distribution model coupled to a correctly specified PK/PD or PBPK model will then result in predictions accurately reflecting the impact of covariates. 
In contrast, if the combination of a covariate distribution model type and one specific PK/PD or PBPK model resulted in good predictions for some observable, this would not mean that they are trustworthy when coupling the covariate distribution model to another PK/PD or PBPK model or when predicting other observables.

In summary, our findings demonstrate that non-Gaussian covariate distribution modelling can be successfully applied to realistic life science covariate datasets.

\backmatter

\bmhead{Acknowledgements}

We thank Laura Zwep and Yushen Guo (University of Leiden, Netherlands) for fruitful discussions on copula modelling, as well as the Mathematical Modelling and Systems Biology group at University of Potsdam, Germany for helpful comments on the manuscript.

\bmhead{Funding statement}

The research of NH has been partially funded by the Deutsche
Forschungsgemeinschaft (DFG) – Project-ID 318763901 – SFB1294.

\begin{appendices}

\section{KL divergence estimation for continuous data}
\label{sec:kld-estimator}

%
To define the bias-reduced nearest-neighbour KL divergence estimator, we denote by $\rho_{k}(i)$ and $\nu_{k}(i)$ the distance to the $k$-th nearest neighbour of $x^{(i)}$ in $\mathbf{x}\setminus x^{(i)}$ and in $\mathbf{y}$, respectively. 
Furthermore, let $\eps_{i} = \max(\rho_{1}(i),\nu_{1}(i))$ be the larger of the distances to the first nearest neighbour of $x^{(i)}$ in $\mathbf{x}$ and $\mathbf{y}$, and $k_{i}, l_{i}$ (one of $k_{i}, l_{i}$ is always 1) be the largest possible values such that $\rho_{k_{i}}(i)  \leq \eps_{i}$ and $\nu_{l_{i}}(i) \leq \eps_{i}$.
Then, the bias-corrected KL divergence estimator by \cite{Wang2009} is given by
\begin{equation}
\label{eq:kld-estimator}
\widehat{D_\text{KL}}(\mathbf{x},\mathbf{y}) = \frac{d}{n}\sum_{i=1}^{n}\log \frac{\nu_{l_{i}}(i)}{\rho_{k_{i}}(i)} + \log \frac{m}{n-1} + \frac{1}{n}\sum_{i=1}^{n}\Big(\psi(l_{i})-\psi(k_{i})\Big).
\end{equation}
The first two summands are a Monte Carlo approximation of KL divergence using nearest neighbour density estimates, the last term (involving the digamma function $\psi(x) = \frac{d}{dx}\log \Gamma(x)$) compensates the asymptotic bias of nearest neighbour density estimation, and the use of different numbers of neighbours in $\rho$ and $\nu$ counterbalances finite sample bias originating from different convergence speed of nearest neighbour density estimates \cite[for details, see][]{Wang2009}.

Since samples obtained by the MICE method may contain duplicates but the nearest neighbour-based KL divergence estimator \eqref{eq:kld-estimator} cannot deal with duplicates in the data, we apply post-processing to the MICE samples by selecting only unique items.

\section{Subsampling for uncertainty quantification of KL divergence estimates}
\label{sec:subsampling}

The following description of the generic subsampling procedure is modified after \cite{Geyer-Notes}.
Let $\hat\theta_{n} = \hat\theta_{n}(x^{(1)},...,x^{(n)})$ be an estimator for a parameter of interest $\theta$ such that $\tau_{n}(\hat\theta_{n} - \theta)$ converges in distribution to some cumulative distribution function $G$. 
Asymptotically, it follows that $\mathbb{P}\big[G^{-1}(\alpha/2) < \tau_{n}(\hat\theta_{n} - \theta) < G^{-1}(1-\alpha/2)\big] \rightarrow 1-\alpha$.
If $\theta^{*}_{b}$ denotes the estimator for a subsample from $\mathbf{x}$ of size $b$ such that $b\rightarrow\infty$ and $b/n\rightarrow 0$ as $n\rightarrow\infty$, then $\tau_{b}(\theta^{*}_{b} - \hat\theta_{n})$ converges in distribution to the same limiting distribution $G$. Also, if $G_{s,b}$ denotes the empirical cumulative distribution function of $s$ subsamples of size $b$, with $s\rightarrow\infty$ as $n\rightarrow\infty$, then $G_{s,b}\rightarrow G$.
Combining these equations, it follows that 
\begin{equation}
\label{eq:ci-subsampling}
\mathbb{P}\big[G_{s,b}^{-1}(\alpha/2) < \tau_{n}(\hat\theta_{n} - \theta) < G^{-1}_{s,b}(1-\alpha/2)\big] \rightarrow 1-\alpha
\end{equation}
as $n\rightarrow\infty$. Hence, by solving the inequality in brackets in \eqref{eq:ci-subsampling} for $\theta$, an asymptotic $(1-\alpha)$ confidence interval for $\theta$ is obtained.

In our setting, $\theta = D_\text{KL}(p||q)$ and $\hat\theta = \widehat{D_\text{KL}}(\mathbf{x},\mathbf{y})$ is the estimator \eqref{eq:kld-estimator}. 

The convergence rate $\tau_{n}$ of this estimator has not been studied so far, 
but results are available for a non-bias-corrected 1-nearest neighbour density estimator. 
Depending on the regularity of $p$ and $q$ as well as the dimension $d$, a rate $\tau_{n} = \sqrt{n}$ can be obtained \cite{Noh2014, Singh2016,Zhao2020}. 
We confirmed this convergence rate empirically for the estimator \eqref{eq:kld-estimator} using the approach described in \cite[Ch.~8]{Politis1999}.

\section{KL divergence estimation for combined continuous/discrete data}
\label{sec:combined-framework}

Here, we describe KL divergence estimation in the general case where some covariates are continuous and others discrete.
Decomposing $x = (x_{c},x_{d})$ into the continuous and discrete parts $x_{c}$ and $x_{d}$, 
respectively, KL divergence between $p$ and $q$ is then given by 

\begin{equation}
\label{eq:KLD-comb}
D_\text{KL}(p||q) := \sum\limits_{x_{d}}\int \log\left(\frac{p(x_{c},x_{d})}{q(x_{c},x_{d})}\right)p(x_{c},x_{d})dx_{c}.
\end{equation}
Conditioning on the discrete variables and reordering, we can rewrite $D_\text{KL}(p||q)$ as
\begin{equation}
\label{eq:KLD-comb-rewrite}
D_\text{KL}(p||q) = \sum\limits_{x_{d}}p(x_{d})\int \log\left(\frac{p(x_{c}|x_{d})}{q(x_{c}|x_{d})}\right)p(x_{c}|x_{d}) dx_{c} + \sum\limits_{x_{d}} \log\left(\frac{p(x_{d})}{q(x_{d})}\right)p(x_{d}),
\end{equation}
known as the chain rule for KL divergence \cite[p.115f]{Schervish1995}.

The integral in the first summand can be estimated on the stratified data $\mathbf{x}_{c}|x_{d}$ and $\mathbf{y}_{c}|x_{d}$ using the estimator \eqref{eq:kld-estimator} for continuous data described in the main text (where $x_{c}\in\mathbf{x}_{c}|x_{d} \Leftrightarrow (x_{c},x_{d})\in\mathbf{x}$). The terms $p(x_{d})$ and $q(x_{d})$ are estimated using relative frequencies of $x_{d}$ in $\mathbf{x}$ and $\mathbf{y}$, respectively; for details, see \cite{kldest-Package}.

\section{Scale dependency of KL divergence estimates}
\label{sec:scale-dependency}

\begin{figure}[htp!]
\centering
\includegraphics[width=\textwidth]{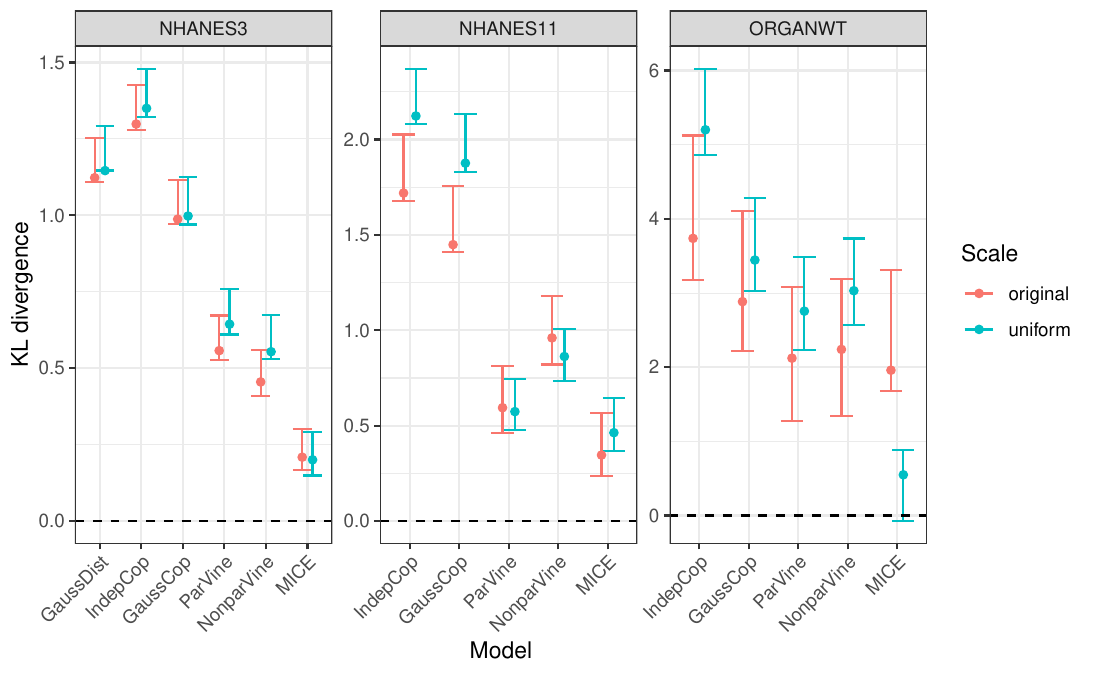}
\caption{
\label{fig:scale-dependency}
KL divergence estimates (with 95\% CI) on original and uniform scale.
}
\end{figure}

While KL divergence is scale-invariant, the estimator \eqref{eq:kld-estimator} might differ between scales. In the main text, all KL divergence estimates are computed on the original data scale.
We investigated a potential scale dependency by computing KL divergence on both the original scale (using samples $\mathbf{x}$ and $\mathbf{y}$) and the uniform scale (using the transformed samples $\hat u(\mathbf{x})$ and $\hat u(\mathbf{y})$, cf.~\eqref{eq:scale-transform}) and found that the impact was very low for NHANES-3, low for NHANES-11 datasets, and at most moderate for the ORGANWT dataset (see Fig.~\ref{fig:scale-dependency}).

\clearpage
\section{Supplementary Figures}
\label{sec:supp-figs}

\begin{figure}[htp!]
\centering
\includegraphics[width=\textwidth]{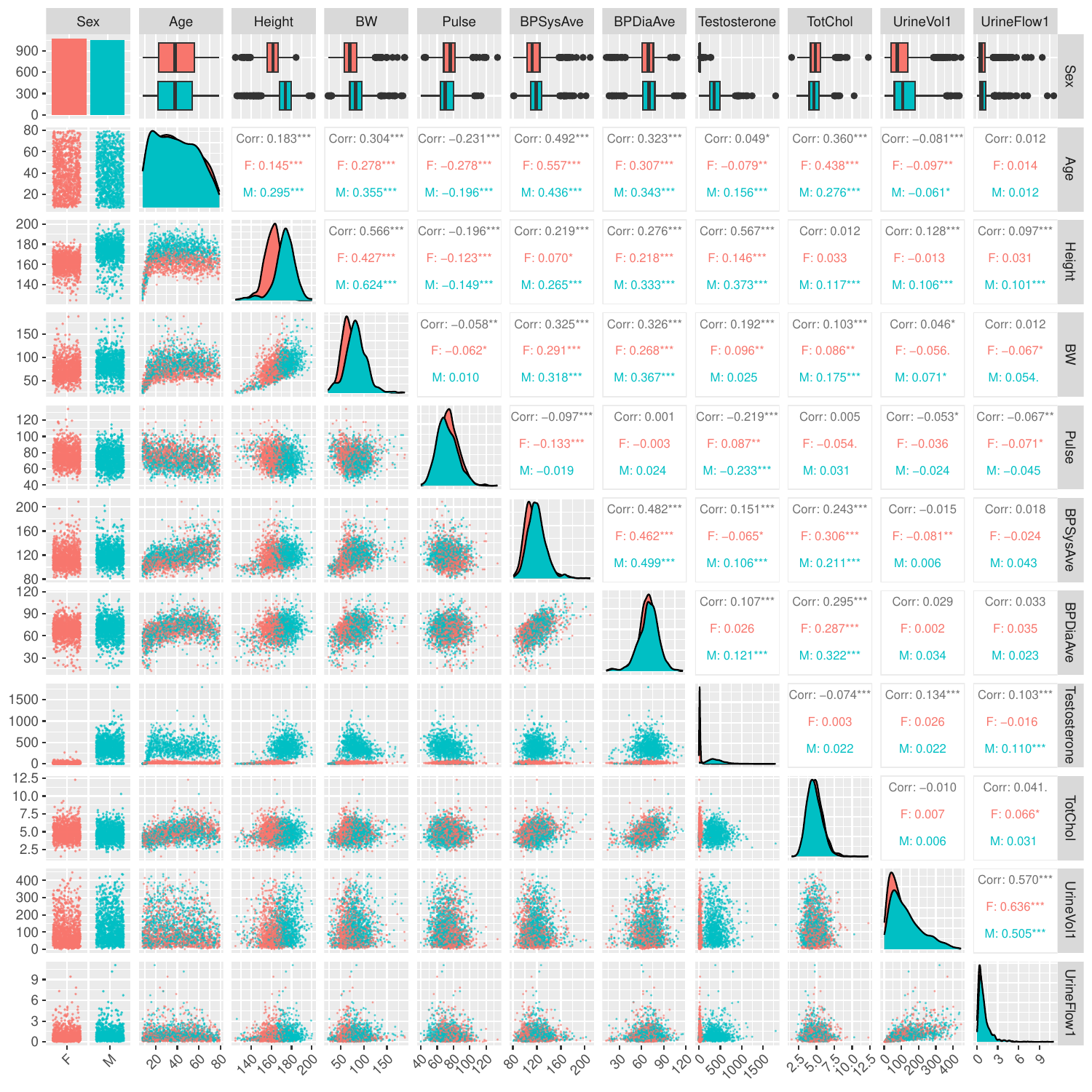}
\caption{
\label{fig:pairs-nhanes11}
Scatter plot matrix of the NHANES-11 data. 
Abbreviations [units]: Age [years]; height [cm]; BW, body weight [kg]; Pulse, heart rate [beats/minute]; BPSysAve, average systolic blood pressure [mmHg]; BPDiaAve, average diastolic blood pressure [mmHg]; Testosterone, testosterone concentrations [ng/dL]; TotChol, total cholesterol [mmol/L]; UrinVol1, urine volume [mL]; UrineFlow1, urine flow rate [mL/min].
}
\end{figure}

\begin{figure}[htp!]
\centering
\includegraphics[width=\textwidth]{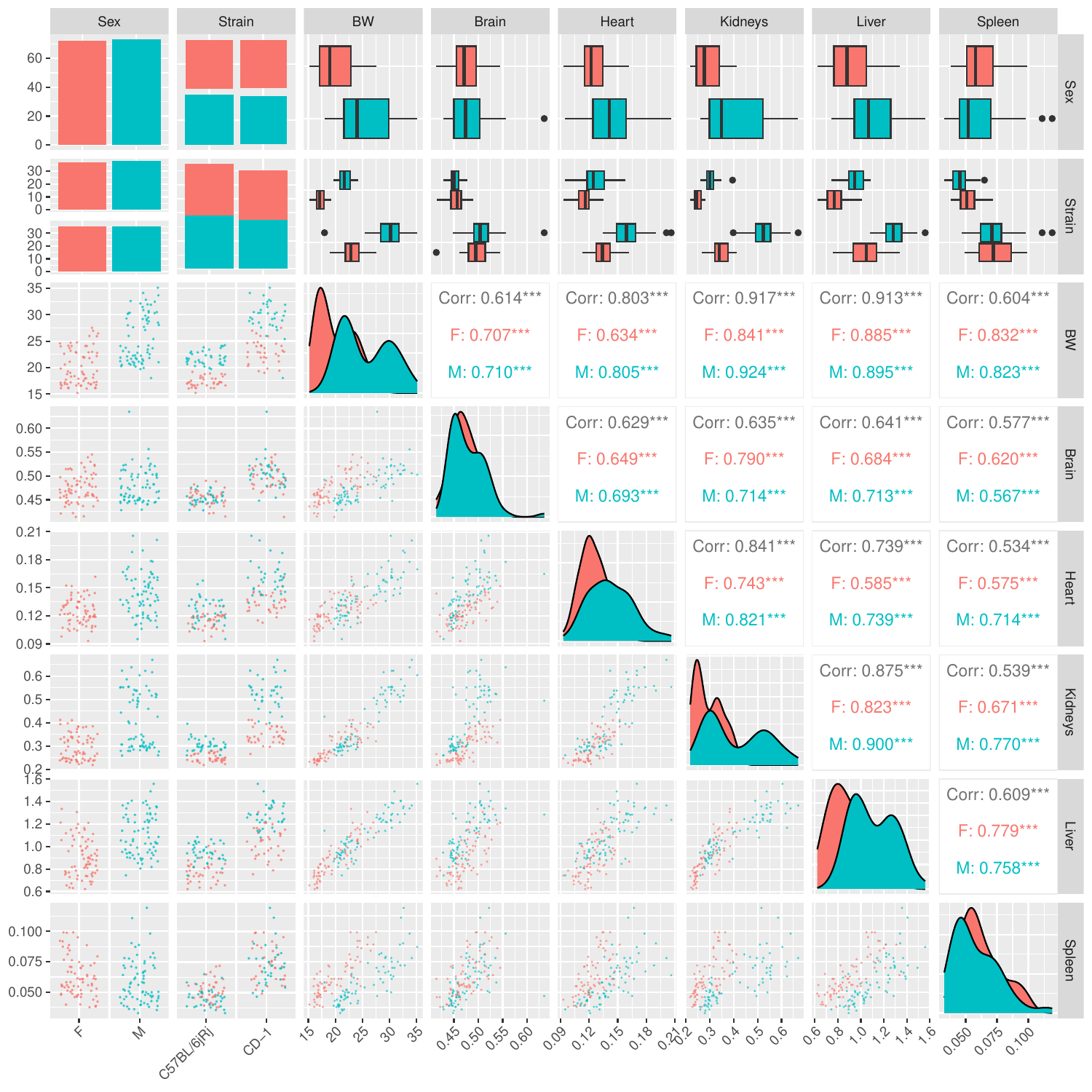}
\caption{
\label{fig:pairs-organwt}
Scatter plot matrix of the ORGANWT data.
Abbreviations [units]: BW, body weight [g]; Brain, brain organ weight [g]; Heart, heart organ weight [g]; Kidney, kidney organ weight [g]; Liver, liver organ weight [g]; Spleen, spleen organ weight [g].
}
\end{figure}


\end{appendices}

\bibliography{Literature-VineCopula-KLD}

\end{document}